\documentstyle[pre,aps,multicol,epsf]{revtex} 
\newcommand{\hz} {H_{0}}
\newcommand{\half} {\frac{1}{2}} 
\newcommand{\vmax} {V_{\rm max}}
\newcommand{\vmin} {V_{\rm min}} 
\begin{document} 
\draft
\title{Pseudo-Boundaries in Discontinuous 2-Dimensional Maps} 
\author{Oded Farago and Yacov Kantor} 
\address{School of Physics and Astronomy, Tel Aviv University, 
Tel Aviv 69 978, Israel} 
\maketitle 
\begin{abstract} 
It is known that Kolmogorov-Arnold-Moser boundaries appear in 
sufficiently smooth 2-dimensional area-preserving maps. When such
boundaries are destroyed, they become pseudo-boundaries. We show that
pseudo-boundaries can also be found in discontinuous maps. The origin 
of these pseudo-boundaries are groups of chains of islands which 
separate parts of the phase space and need to be crossed in order to 
move between the different sub-spaces. Trajectories, however, do not
easily cross these chains, but tend to propagate along them. This type
of behavior is demonstrated using a ``generalized'' Fermi map.  
\end{abstract}
\begin{multicols}{2} 
\narrowtext

The most obvious way to set particles into motion is to subject them to an
external field. Yet, particle currents can be also induced by zero
averaged time-dependent forces, provided an asymmetric potential is
applied in the system. Recent studies of systems consisting of Brownian
particles in periodic, locally asymmetric, potentials have shown that
motion may be induced by periodic forcing, non-random noises or due to an
oscillatory change in the potential profile \cite{direct}. Such
rectification processes are interesting because they can lead to the
generation of new microscopic pumping techniques. They may also provide
some insight into the mechanism of protein motors \cite{protein}. 
Another example for a net directed motion of particles in periodic
(both in time and space) potentials is found in a model describing the
motion of a classical particle within an infinite set of equally
spaced potential barriers, whose position and height oscillate
periodically \cite{Jaric}. At each collision, the particle either
crosses or reflects from the barrier, depending whether or not its
kinetic energy is larger than the potential height. The special
feature of this system (unlike these in Refs.~\cite{direct}) is its
chaotic deterministic nature. Chaotic dynamics are frequently studied
by investigating the corresponding trajectories in the system's phase
space. Our system is one-dimensional with a time dependent
Hamiltonian, hence it's phase space is four-dimensional. The
Poincar\`{e} method can be applied to reduce this 4-dimensional phase
space into a 2-dimensional one by taking a surface lying in the phase
space and studying the sets of intersections of different trajectories
with this surface. Successive crossing points are computed using a
2-dimensional area-preserving map of the surface, the reduced phase
space, onto itself. In order to inquire the features of this map, we
constructed a picture of it's 2-dimensional phase space by plotting
the sets of points (which are the trajectories in the reduced phase
space) starting at different initial conditions \cite{KanJar,Farago}.

\begin{figure}
\epsfysize=14.5\baselineskip
\centerline{\hbox{
      \epsffile{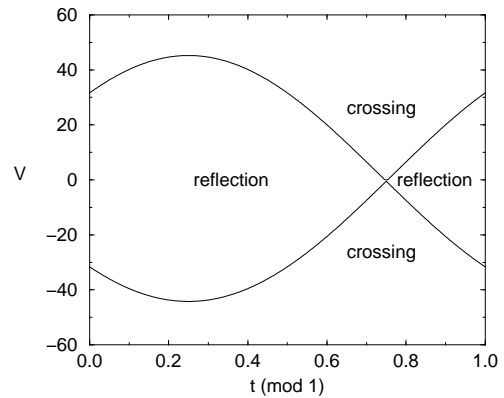}  }} 
\caption   {\protect\footnotesize Regions in the $(t-V)$ phase space
           of points mapped by the sub-maps of crossing and
           reflection, for a barrier mean height $\hz=500$. The map is
           discontinuous over the curves separating between the
           different regions.}
\label{refcross}
\end{figure}

In this paper we report and qualitatively explain an interesting
phenomenon observed while studying this map. Pictures of the phase
space, constructed for different values of the map's parameters,
displayed that the phase space consists of a {\em stochastic sea},
which a typical trajectory covers ergodically and some embedded
islands, in which the motion is mainly {\em regular}\/ over closed,
elliptic-like, curves. This structure is quite common to many
area-preserving maps. We did not observe Kolmogorov-Arnold-Moser
(KAM) curves in our phase space. These curves of regular motion, if
they exist, bound different parts of the stochastic sea, that is,
prevent trajectories from leaving or entering them. According to the
KAM theorem \cite{Kolmo,Arnold,Moser}, KAM curves do not appear in
discontinuous maps as is ours~\cite{smooth}. The origin of
discontinuity in our map can be easily understood:~Points in the phase
space are divided into two classes each corresponding to either
crossing or reflection from barriers. For each class the map is
represented by one of its two possible ``sub-maps''. Discontinuity
appears on the boundaries between regions containing points of
different classes (see Fig.\ \ref{refcross}). 

Since KAM curves do not exist, the stochastic sea is a single
connected unit with only a few islands excluded~\cite{island}. The
motion over a connected (invariant) component of the phase space is
ergodic and therefore a trajectory starting at almost any point in the
stochastic sea, will eventually reveal this entire region. Numerical
observations have shown that the number of iterations needed to
explore completely the stochastic sea varies, within more than 3
orders of magnitude, for different values of the map's parameters. In
the cases when this number was considerably large, the stochastic sea
seemed to be divided into smaller parts in which trajectories traveled
for long time intervals before moving from the one to the other. The
boundaries between the different parts are called {\em pseudo}\/ ({\em
  partial}\/) boundaries. Their appearance does not violate the
ergodicity hypothesis, it simply implies that it is valid only for
time scales much larger than the characteristic escape time from each
pseudo-bounded part of the stochastic sea.     

Each KAM boundary is destroyed when the map's control parameter
exceeds some critical value. At the critical point an infinite number
of gaps are formed in the torus and it turns into a Cantor set (hence
called a {\em cantorus})~\cite{Aub,Per}. As we continue to change the
control parameter, the flux across the cantorus increases from its
zero value at the critical point \cite{flux}. Thus, close to the
critical point, the flux across a cantorus is low and cantorus serves
as a  pseudo-boundary. Cantori, like KAM tori, do not appear in
discontinuous maps. We show in this paper that pseudo-boundaries are
found in discontinuous maps due to other reasons. They are formed by
chains of islands surrounding different areas of the phase space. In
order to leave these areas, trajectories need to diffuse {\em across}
the chains, which usually appear in groups. However, the motion in the
vicinity of the chains, tends to be {\em along} the chains, in a way
that resembles the regular motion over the chains.   
 
In smooth maps, cantori and chains of islands are deeply related. For
each torus, cantorus or chain of islands, there is a corresponding
{\em winding number} (for definition of winding numbers see, for example,
\cite{LibLicbk}). Tori and cantori have irrational winding numbers
while chains of islands have rational ones. Green \cite{green} has
shown that close to any torus or cantorus, there are infinitely many
chains of islands whose winding numbers are successive finite
truncations of the continued fraction representing the irrational
winding number of the torus or cantorus. It has also been shown
\cite{Macetal,BenKad} that the flux across a cantorus is the limit of
the flux across its approximating chains of islands. This last result
means that chains of islands which approximate cantori that serve as
pseudo-boundaries (i.e., which the flux across them is low) are
pseudo-boundaries themselves. For discontinuous maps, in which winding
numbers are meaningless, the properties of the chains should be
studied differently, since they appear with no relation to cantori. A
special emphasis should be given to the way in which trajectories
behave when they are trapped between several adjacent chains.

We consider a classical particle of unit mass, moving within
a one-dimensional sequence of equidistant potential barriers, each
of an infinitesimal width and finite height, $H(t)$. The barriers
oscillate with the same frequency and in phase. Their velocity is
given by $v(t)=v_{b}g(t)$, while $H(t)=\hz[1+g(t)]$, where
$g(t)=\sin(2\pi t)$. The particle moves freely between
the barriers, occasionally colliding with them. At each impact it can
either cross or be reflected from the barrier. It crosses the barrier
if its kinetic energy in the reference frame of the barrier exceeds
the height of the potential barrier at the moment of impact,
i.e., $|V_{n}-v(t_{n})|>\sqrt{2H(t_{n})}$, where $t_{n}$ and $V_{n}$
are, respectively, the time of the $n$-th impact and the velocity of
the particle before that impact. If the particle crosses the barrier,
its velocity is not changed, i.e., $V_{n+1}=V_{n}$. If it is
reflected, on the other hand, it acquires a new velocity:
$V_{n+1}=-V_{n}+2v(t_{n})$. In order to calculate the moment of the
next impact, one needs to consider the motion of both the particle and
the barriers. We made the approximation that the distance traveled
between two consecutive collisions is constant and equals to the
distance between to barriers, $L$, as if the barriers do not change
their position. We also assumed that in a case of reflection, the
particle is reflected backward and therefore if $V_{n+1}$ has the same
sign as $V_{n}$, we replace it with $-V_{n+1}$. A similar
approach was used in Ref.\ \cite{LibLicpr} for the Fermi accelerator
model (a ball bouncing between two walls \cite{Fermi}). Due to the
periodicity of $g(t)$ only time modulo period (=1), which we denote as
phase, is relevant, and two consecutive phases are related by
$t_{n+1}=t_{n}+L\,/\,|V_{n+1}|\ \mbox{(mod 1)}$.   

\begin{figure}
\epsfysize=14\baselineskip
\centerline{\hbox{
      \epsffile{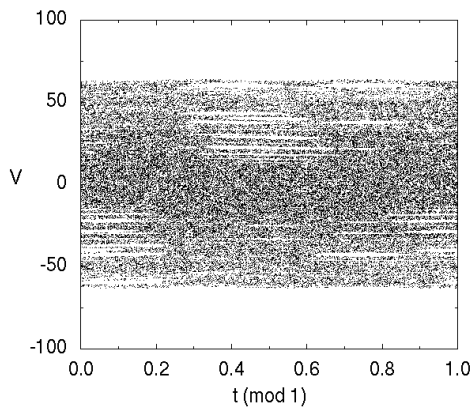}  }}
\caption   {\protect\footnotesize The subspace explored after
            $N=2\cdot10^{10}$ iterations of the map for a barrier mean
            height $\hz=2500$. The points denote the velocity of the
            particle $V_{n}$ (y-axis), and the phase of the barrier,
            $t_{n}$ (x-axis), at the $n$-th ($n=1,2,\ldots,N$)
            collision.}
\label{pkb1}
\end{figure}

We chose the mean height of a barrier, $\hz $, to be the control
parameter, while we set the other parameters $L=100$ and $v_{b}=\half
$. For the special case $\hz=\infty $, barrier crossing is impossible
and the map reduces to the Fermi map \cite{fermap}. For finite $\hz $,
the map is discontinuous. Fig.\ \ref{refcross} shows the discontinuity
lines in the $(V-t)$ phase space which separate the regions mapped by
the different sub-maps of crossing and reflection (for $\hz=500$). The
map is area-preserving since both sub-maps are area-preserving and
since their ranges do not overlap. For $V>\vmax\equiv
2\sqrt{\hz}+v_{b}$ and $V<\vmin\equiv -2\sqrt{\hz}+v_{b}$, points in
the $(V-t)$ phase space are mapped only by the crossing sub-map (the
kinetic energy is sufficiently large to cross the barriers at any
phase), hence the motion in this part of the phase space is over the
lines $V=V_{0}=\mbox{const}$. The stochastic sea lies between $\vmin
$ and $\vmax $, including several islands. As already explained, a
picture of it is made by starting at almost any one of its points,
iterating the map a sufficiently large number of times and plotting
the resulting trajectories, $\{(t_{n},V_{n});n=0,1,2\ldots\}$.
We performed this procedure for various values of $\hz $. For $\hz $
values below 1000, the stochastic sea was completely explored up
to the resolution of the plots) after $10^{6}-10^{7}$ iterations. As
expected, apart from some embedded chains of islands, the trajectory
wandered in the entire region between $\vmin $ and $\vmax $. For
$\hz >1000$, the situation was quite different. Fig.\ \ref{pkb1}
depicts the velocities $V_{n}$ and phases $t_{n}$ of first $2\cdot
10^{10}$ collisions between the particle and the barriers. The
velocities appear to remain smaller than 63, while the stochastic sea
ranges between $\vmin =-99.5$ and $\vmax =100.5$.

\begin{figure}
\epsfysize=17\baselineskip
\centerline{\hbox{
      \epsffile{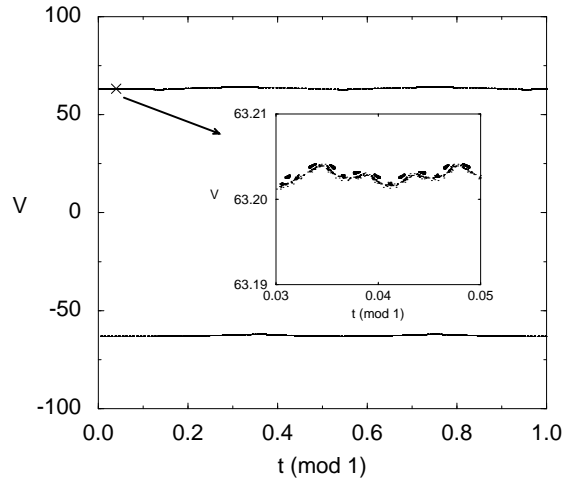}  }}
\caption  {\protect\footnotesize A chain which appears in phase space
           around $|V|\sim 63$ for a barrier mean height
           $\hz=2500$. Its shape resembles the boundary of the area
           shown in Fig.\ \protect\ref{pkb1}. The inset is a
           magnification of a small part of the chain, showing few of
           its islands. A narrow stochastic layer is located just
           below the chain which serves as a
           pseudo-boundary. Trajectories that reaches the stochastic
           layer, may move over it for thousands of iterations before
           either falling back into the stochastic sea (the usual
           case) or crossing the chain (the rare scenario).}
\label{pkb2}
\end{figure}

\begin{figure}
\epsfysize=17\baselineskip
\centerline{\hbox{
      \epsffile{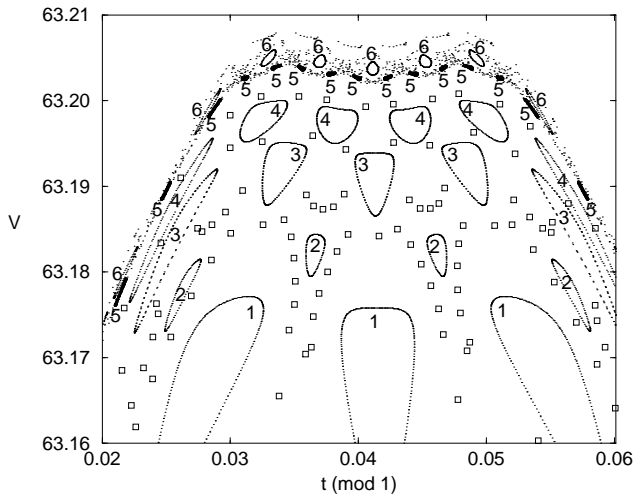}  }}
\caption  {\protect\footnotesize A few adjacent chains of islands
           found in the phase space for a barrier mean height
           $\hz=2500$, forming together a pseudo-boundary. The numbers
           inside the islands label the different chains. The number
           of islands in the chains labeled 1 to 6, in the upper
           (lower) half of the phase space, are respectively: 27
           (33), 37 (47), 49 (59), 60 (72), 420 (466), 157 (164). The
           dots and the open squares belong to two stochastic
           trajectories wandering, respectively, above and below the
           chain labeled 5. (Only a small part of the positive
           velocity branches of the chains is shown here.)}
\label{pkb3}
\end{figure}

A pseudo-boundary exists around $|V|\sim63 $. A closer look of
that area of the phase space reveals a few adjacent chains of islands,
located one above the other, whose shapes are similar to the boundary
of the area plotted in Fig.\ \ref{pkb1}. One of these chains is shown
in Fig.~\ref{pkb2}. It is~composed of two velocity branches, in the
upper ($V>0$)~and the lower ($V<0$) halves of the phase space, between
which the motion alternates. The effect of this chain on the motion
close to it is depicted in the inset of Fig.\ \ref{pkb2}, showing an
enlargement of a small segment of the chain. A narrow stochastic layer
appears below the chain (above it, if the negative velocity branch is
considered). We found that trajectories can move over this narrow
layer for thousands of iterations before leaving it, usually falling
back into the stochastic sea. Two features resemble the behavior close
to a KAM torus or a cantorus: (1)~the motion is restricted to one side
of the boundary (the chain), and (2)~the motion is correlated to the
boundary and only gradually deviates from it. As for a cantorus, if
the chain is long and narrow with only small gaps between the islands,
there are only small areas from which a trajectory can jump from one
of its sides to the other.
Fig.\ \ref{pkb2} demonstrates the effect of a single long chain with
islands located close to each other. The stochastic motion near such
a chain is highly correlated. It, roughly, follows the regular
motion over the chain and only slowly deviates from it. But when a
pseudo-boundary appears, we usually observe a group of several chains,
lying closely one above the other, and not only a single
chain. Trajectories need to diffuse across all these chains in order
to move to another sub-space. However, due to the character of the
motion near the chains, they tend to move along them. Suppose a
trajectory enters between a few chains, close to one of
them. It propagates along the chain, while gradually moving away. If
there are a few nearby chains, than while it moves away from one chain
it gets closer to another, along which it propagates for subsequent
iterations. This is an over-simplified picture since it is not always
possible to relate the motion to a particular chain at each instance,
yet, it qualitatively explains the character of the motion in this
``band of chains''. The trajectory can be trapped between the chains
for many iterations. If it succeeds in crossing all of them, it
escapes. Usually, however, after spending some time wandering between
the chains, it ``falls back'' into the stochastic sea. Fig.~\ref{pkb3}
depicts six chains and two stochastic trajectories. Note that while
one of the trajectories (open squares) travels between the chains
labeled 1 to 5, the other trajectory (dots) is restricted to the area
above the chain labeled~5. Indeed, since the flux across this chain is
very small, compared to the flux across the other chains (probably
because of the fact that it consists of relatively many islands with
small gaps between them), the chain labeled~5 (also shown in the inset
of Fig.\ \ref{pkb2}) serves as the natural limit for both sides of the
pseudo-boundary.     
In conclusion, we have shown that long and narrow chains of islands 
may serve as pseudo-boundaries in discontinuous 2-dimensional
maps. If a group of several adjacent chains surround a certain
sub-space, they all need to be crossed in order to move to a
different sub-space. However, in the vicinity of the chains,
trajectories which stochastically wander between the islands tend to
propagate along the chain, slowly moving from one chain to the
other. Hence, the chains are not easily crossed and they form a
pseudo-boundary.

Acknowledgments: The authors would like to thank to M.\ V.\ Jaric for
useful discussions. This work was supported by the Sackler Institute
for Solid State Physics at the Tel-Aviv University and by the Israel
Science Foundation Grant No.~246/96-1.    

\end{multicols}

\begin{references}
\bibitem{smooth}
   For 2-dimensional nearly integrable maps, it appears that the
   continuity of the first derivative is a necessary condition for the
   existence of KAM curves, while the continuity of the second
   derivative is a sufficient one.
\bibitem{island}
   These elliptic curves are also KAM curves. In the area of phase
   space where they exist, the map is locally smooth.
\bibitem{flux}
   The flux across a curve is the fraction of trajectories (uniformly
   distributed on one of the curve's sides) which cross the curve
   in a time unit (e.g. per iteration). This flux is proportional to
   the total area from which trajectories are mapped form one side of
   the curve to the other.
\bibitem{fermap}   
   The standard Fermi map describes a ball bouncing between a fixed
   and an oscillating wall, while for $\hz=\infty $ our map describes
   a motion between two oscillating walls.
\end{references}
\end{document}